\documentclass[10pt,conference,twocolumn]{IEEEtran}

\usepackage{tikz}
\usepackage{eso-pic}

\AddToShipoutPictureBG*{%
  \AtPageUpperLeft{%
    \put(0,-35){% Adjust the vertical offset (-35) if it overlaps or is too high
      \begin{minipage}{\paperwidth}
        \centering
        \small
        \color{gray} % Optional: make the text subtle
        © 2025 IEEE. Published in IEEE OCEANS 2025 conference. DOI: 10.23919/OCEANS59106.2025.11244960. \\
      \end{minipage}%
    }%
  }%
}

\usepackage{cite}
\usepackage{times}
\usepackage{algpseudocode}
\usepackage{mathtools}
\usepackage{amsmath}
\usepackage{booktabs}
\usepackage{graphicx}
\usepackage[figuresright]{rotating}
\usepackage{pict2e}
\usepackage{algorithm}
\usepackage{amssymb,amsmath,amsthm}
\usepackage{amsmath}
\usepackage{setspace}
\usepackage{comment}
\usepackage{balance}
\usepackage{rotating}
\usepackage{multirow}
\usepackage{wrapfig}
\usepackage{mathtools}
\usepackage{booktabs}
\usepackage{gensymb}
\usepackage{array}
\usepackage[font=footnotesize]{caption}
\usepackage{multicol}
\usepackage{algpseudocode,algorithm,algorithmicx}
\usepackage{colortbl}

\renewcommand{\theequation}{\arabic{equation}}
\usepackage{bm}
\graphicspath{ {./Fig/} }

\begin{document}
%title for journal:
%\title{Terrestrial and Water Body Internet of Things: A  Semantic-Based Self-Learning Approach %~(TWB IoT) }
\title{Intelligent Code-Division Multiplexing for Resilient Underwater Optical Wireless Communications} %~(TWB IoT) 

%old title
%\title{Underwater Internet of Things Interoperability: A Semantic-based Approach}
%
%\author{Ananya~Hazarika,~\IEEEmembership{Student Member,~IEEE} and Mehdi~Rahmati,~\IEEEmembership{Senior Member,~IEEE,}
        
%\thanks{\IEEEcompsocthanksitem A.~Hazarika and M.~Rahmati are with the Department of Electrical Engineering and Computer Science, Cleveland State University, OH, 44115 USA \protect 
%E-mails: 
%a.hazarika@vikes.csuohio.edu and m.rahmati@csuohio.edu.}
%}

\author{{Gaurav Mahadik, Ananya Hazarika, and Mehdi Rahmati}\\ Department of Electrical and Computer Engineering \\ Cleveland State University, OH, USA 44115 \\ Emails: {\{g.mahadik, a.hazarika\}@vikes.csuohio.edu and m.rahmati@csuohio.edu}}
  
%\markboth{Submitted to IEEE Sensors Journal, December~2021}%
%{Shell \MakeLowercase{\textit{et al.}}: Bare Demo of IEEEtran.cls for Journals}

\maketitle
\thispagestyle{empty}
%\thispagestyle{plain}
%\pagestyle{plain}

% As a general rule, do not put math, special symbols or citations
% in the abstract or keywords.
\begin{abstract}
This paper presents a novel intelligent chaotic-based code-division multiple access (CDMA) scheme for underwater optical wireless communication (UOWC), addressing critical performance degradation caused by severe scattering and multipath dispersion in underwater environments. Unlike conventional modulation techniques such as on-off keying, which depend on precise pulse timing and show high sensitivity to channel distortions, the proposed approach leverages 
unpredictable deterministic chaotic sequences generated by the logistic map to enhance robustness against scattering-induced impairments. A Multi-Agent Reinforcement Learning (MARL) framework enables distributed agents to dynamically adapt chaotic map parameters, including initial conditions and bifurcation parameters, based on real-time environmental feedback, optimizing sequence generation to maintain low cross-correlation properties and improve resilience to multipath effects. Experimental validation using a 2-meter water tank testbed with controlled turbidity demonstrates superior performance compared to conventional schemes. The adaptive framework exhibits rapid convergence and relaxed synchronization requirements, making it highly suitable for distributed underwater networks where centralized coordination is impractical.
\end{abstract}


\IEEEpeerreviewmaketitle
\section{Introduction}
Underwater Optical Wireless Communication (UOWC) has emerged as a promising technology for enabling high-speed, low-latency data transmission over short to moderate distances in aquatic environments. This capability makes it particularly suitable for a wide range of applications, including real-time underwater data collection, Autonomous Underwater Vehicle (AUV) coordination, and maritime surveillance. Recent advancements have further expanded its potential; for example, integration with Multiple-Input Multiple-Output~(MIMO) systems and advanced coding has pushed error-free throughput to distances exceeding 100 meters in deep ocean, while hybrid optical-acoustic approaches have improved reliability in turbid conditions~\cite{HAN20191}. Compared to traditional underwater acoustic communication, UOWC offers several advantages, including higher bandwidth, operation in interference-free frequency bands, and moderate transmission loss in water~\cite{fu2022anti}. Intensity Modulation with Direct Detection (IM/DD) is commonly used at the physical layer due to its simplicity and compatibility with LED and laser diode sources\cite{chaaban2021capacity}. Among modulation schemes, On–Off Keying (OOK) is widely adopted for its low complexity, energy efficiency, and cost-effective hardware implementation~\cite{elganimi2013performance}.

A major performance bottleneck in UOWC arises from the scattering in turbid water, which introduces Inter-Symbol Interference~(ISI) and path-dependent fadings. On the other hand, modulation schemes such as OOK suffer from this performance degradation in heavily scattered environments~\cite{10337231}. Therefore, the scalability of OOK-based systems is severely limited, and their performance is tightly coupled to the underwater optical channel's conditions, clarity, and stability of water. While traditional Code Division Multiple Access~(CDMA) codes, i.e., Walsh-Hadamard or Gold sequences, offer orthogonality under ideal synchronized, distortion-free conditions, they may lose their effectiveness in scenarios in which propagation-induced delays disturb code alignment and change cross-correlation properties of the code.
Chaotic codes are dependent on the initial conditions; therefore, by choosing appropriate initial values, they can produce an infinite set of uncorrelated sequences. One widely studied chaotic set that has been employed for underwater acoustic communications~\cite{azou2002chaotic} is generated based on the Logistic map, which produces a variety of distinct sequences
for different users, by changing the initial states and/or
its bifurcation parameter~\cite{rahmati2019network}. 
%~\cite{heidari1994chaotic}.

We propose an adaptive chaotic CDMA code generator based on Reinforcement Learning~(RL). Unlike fixed-seed chaotic sequences, our proposed approach enables real-time adaptation of the chaotic system's initial condition and parameters for multiple users, considering channel conditions and a learned system performance objective. Chaotic sequences~\cite{broomhead1999codes}, generated from nonlinear dynamical systems, provide an alternative to conventional spreading techniques. Unlike periodic or pseudo-random sequences, deterministic chaotic codes show acceptable spectral characteristics and maintain low cross-correlation properties even under time shifts. Our contributions are as follows.

\begin{itemize}   
\item We enable simultaneous transmission using optical chaotic CDMA codes, which are easily produced locally using a secret seed. The approach supports a flexible and large code family, enhancing multiuser access.
\item The proposed chaotic-based CDMA scheme maintains low cross-correlation between users, improving robustness against scattering-induced multipath distortion in underwater optical channels. Additionally, it exhibits relaxed synchronization requirements compared to traditional orthogonal codes, making it more scalable for distributed underwater networks.
%\item Spread spectrum improves the resiliency against spectral nulls caused by water turbidity.
\item  A Multi-Agent Reinforcement Learning (MARL) framework is employed to dynamically adapt the initial conditions and bifurcation parameters of the chaotic logistic maps across users, enabling coordinated and environment-aware optimization of spreading sequences in underwater optical channels.
\end{itemize}

\section{Related Work}
Underwater CDMA has been studied in both acoustic and optical~\cite{7593257} domains to support multiuser access, mitigate multipath fading, and enhance link reliability~\cite{chung1989optical}. Authors in~\cite{liu2019chaotic} developed a Multi-Carrier CDMA~(MC‐CDMA) employing chaotic sequences to suppress Peak-to-Average Power Ratio~(PAPR). Authors in~\cite{akhoundi2017underwater} implemented a time‐spread optical orthogonal CDMA positioning system, achieving sub‐meter accuracy while sustaining 10 Mb/s per user, exploiting code autocorrelation sidelobe suppression to decouple ranging from data streams. 
More recently, adaptive and machine‐learning‐driven underwater schemes have emerged. Authors in~\cite{lin2024deep} proposed a Gated Recurrent Unit~(GRU)‐based decoder for code‐index cyclic shift keying spread spectrum, which learns temporal channel statistics in shallow water scenarios. 
Spreading sequences in spread spectrum systems are deterministic binary or multi-level patterns with key properties including low auto-correlation for synchronization and low cross-correlation for minimizing multi-user interference. Commonly used sequences are maximal-length (m-sequences) and Gold sequences, which offer a good balance between length and correlation properties. Sequence design significantly impacts system capacity, error performance, and security in underwater optical CDMA~\cite{chung1989optical}.
Chaotic sequences have emerged as an alternative to conventional spreading techniques, potentially offering unique advantages for underwater communication systems. These waveforms are generated by nonlinear dynamical systems exhibiting sensitive dependence on initial conditions. They provide aperiodic, broadband signals with excellent correlation properties and enhanced security features. In chaotic CDMA systems, chaotic sequences enhance privacy and multi-user separation in hostile underwater environments~\cite{ott2002chaos}. Previous work in underwater acoustic communications has demonstrated the effectiveness of chaotic sequences, particularly those generated using the Logistic map~\cite{azou2002chaotic,rahmati2019network}. 
%
%Common generators of chaotic codes include the logistic and tent maps, which produce sequences with uniform statistical distribution. Chaotic maps are discrete-time nonlinear transformations that generate complex dynamics from simple equations, with examples including the logistic map and the H\'enon map~\cite{al2025chaotic}. Map parameters control the degree of chaos, sequence length, and statistical properties. Proper parameter selection ensures low correlation and high unpredictability for underwater optical CDMA~\cite{ott2002chaos}.
%
Unlike traditional CDMA codes such as Walsh-Hadamard or Gold sequences that offer orthogonality under ideal synchronized, distortion-free conditions, chaotic codes maintain their effectiveness even when propagation-induced delays disturb code alignment and degrade cross-correlation properties. This robustness makes chaotic sequences particularly suitable for underwater optical channels, where scattering and multipath effects create challenging propagation conditions that can severely degrade the performance of conventional orthogonal codes.

\section{Proposed Solution}\label{sec:sol}

%This section presents our intelligent chaotic CDMA scheme for UOWC, incorporating a MARL framework for adaptive parameter optimization.
%\subsection{Chaotic Sequence Generation}
 Chaotic sequences are generated by nonlinear dynamical systems that exhibit sensitive dependence on initial conditions. They provide aperiodic, broadband signals with excellent correlation properties and high security. 
Our approach utilizes a logistic map as the fundamental chaotic system for generating spreading sequences. The logistic map is defined as \begin{equation} x_{n+1} = r x_n (1 - x_n), 
\end{equation} 
where $x_n \in [0,1]$ represents the state at iteration $n$, $r \in [0,4]$ is the bifurcation parameter controlling the chaotic behavior, and $x_0$ is the initial condition or seed value. 
%For chaotic behavior, we constrain $r \in [3.57, 4]$, ensuring the system exhibits sensitive dependence on initial conditions and aperiodic behavior.
The continuous chaotic output from the logistic map must be transformed into discrete binary sequences suitable for CDMA spreading operations. This conversion process critically influences the correlation properties and spectral characteristics of the resulting spreading codes. The fundamental thresholding operation is mathematically expressed as 
\begin{equation} 
c_n^{(k)} = 
\begin{cases} 1 & \text{if } x_n^{(k)} \geq \lambda^{(k)} \\ 0 & \text{if } x_n^{(k)} < \lambda^{(k)}, \end{cases} 
\end{equation} 
where $c_n^{(k)}$ represents the $n$-th chip of the spreading sequence for user $k$, $x_n^{(k)}$ is the chaotic state for user $k$ at iteration $n$, and $\lambda^{(k)}$ is the user-specific threshold parameter. While the conventional choice sets $\lambda^{(k)} = 0.5$ to exploit the uniform distribution property of chaotic sequences generated by the logistic map in its chaotic regime, our adaptive framework allows for dynamic threshold adjustment to optimize sequence properties.
The binary sequence generation process extends over $N_c$ iterations, where $N_c$ represents the spreading factor or processing gain of the CDMA system. The complete spreading sequence for user $k$ is therefore defined as the vector $\mathbf{c}^{(k)} = [c_0^{(k)}, c_1^{(k)}, \ldots, c_{N_c-1}^{(k)}]^T$. 
%The choice of $N_c$ directly impacts both the system's multiple access capability and its resistance to multipath interference in the underwater optical channel.
%
The binary sequences are subsequently converted to bipolar format through the transformation,
\begin{equation} 
s_n^{(k)} = 2c_n^{(k)} - 1 = \begin{cases} +1 & \text{if } c_n^{(k)} = 1 \\ -1 & \text{if } c_n^{(k)} = 0 
\end{cases} 
\end{equation}
This bipolar representation $\mathbf{s}^{(k)} = [s_0^{(k)}, s_1^{(k)}, \ldots, s_{N_c-1}^{(k)}]^T$ facilitates direct multiplication with the data symbols and enables efficient correlation-based detection at the receiver. The cross-correlation function between sequences of users $i$ and $j$ is defined as \begin{equation} R_{ij}(\tau) = \frac{1}{N} \sum_{n=0}^{N-1} c_n^{(i)} \cdot c_{(n+\tau) \bmod N}^{(j)} \end{equation} where $N$ is the sequence length and $\tau$ is the time shift. For effective CDMA operation, we require: \(|R_{ij}(\tau)| \leq \theta, \quad \forall i \neq j, \forall \tau \), where $\theta$ is a small threshold value, typically constrained such that $\theta < 0.1$. Note that we need to shift the bipolar sequence by adding a constant bias since the optical intensity should be positive. At the receiver, we subtract the bias before correlation.

%\subsection{Multi-Agent Reinforcement Learning Framework}

%
We implement a MARL system where each agent $A_k$ corresponds to a transmitter generating chaotic sequences for user $k$. The system supports $K$ concurrent users, with each agent operating independently while coordinating through the shared environment. The distributed nature of this architecture enables scalable deployment in underwater networks where centralized coordination may be impractical due to communication constraints and dynamic topology changes.

\textbf{Agent Architecture:} Each agent manages the generation of chaotic sequences for its assigned user by adjusting the parameters of the logistic map. The agents interact with the underwater optical channel, characterized by distortion, scattering, and noise. By learning from environmental feedback, such as bit error rate (BER), signal-to-noise ratio (SNR), and inter-user interference, the agents adapt their sequence parameters to optimize performance. The decentralized nature of the MARL system allows each agent to make local decisions while contributing to the global goal of minimizing interference and maximizing throughput across the network.

\textbf{State Space Definition:} The state of each agent captures both local chaotic system parameters and global network performance metrics. The state comprises the current initial condition of the logistic map, which sets the starting point of the chaotic sequence, and the bifurcation parameter, which governs the chaotic behavior. Additionally, it includes the measured BER to assess transmission reliability, the SNR to evaluate signal quality, the delay spread variance to account for multipath-induced temporal dispersion, and the cross-correlation vector with other users to measure inter-user interference. This comprehensive state representation enables agents to balance local sequence generation with network-wide performance objectives. 

\textbf{Action Space:}
Each agent can adjust its chaotic map parameters to optimize sequence generations. The actions consist of small adjustments to the initial condition of the logistic map, constrained within $[0.001, 0.999]$ to ensure valid chaotic behavior, and incremental changes to the bifurcation parameter, limited to $[3.57, 4.0]$ to maintain the chaotic regime. These bounded adjustments enable agents to explore the parameter space effectively while ensuring that the generated sequences remain chaotic and suitable for CDMA operations. The constrained action space prevents the system from drifting into non-chaotic regimes, which could degrade performance.

\textbf{Reward Function:} The reward function is designed to balance multiple objectives, encouraging agents to minimize error, and encouraging agents to generate sequences that minimize errors, reduce inter-user interference, and maximize data-throughput. 
This multi-objective reward structure encourages agents to collaboratively adapt their chaotic sequences, achieving robust performance in the presence of scattering and multipath effects.

\section{Results and Discussions}

\begin{figure*}[!t]
    \centering
\includegraphics[width=\linewidth]{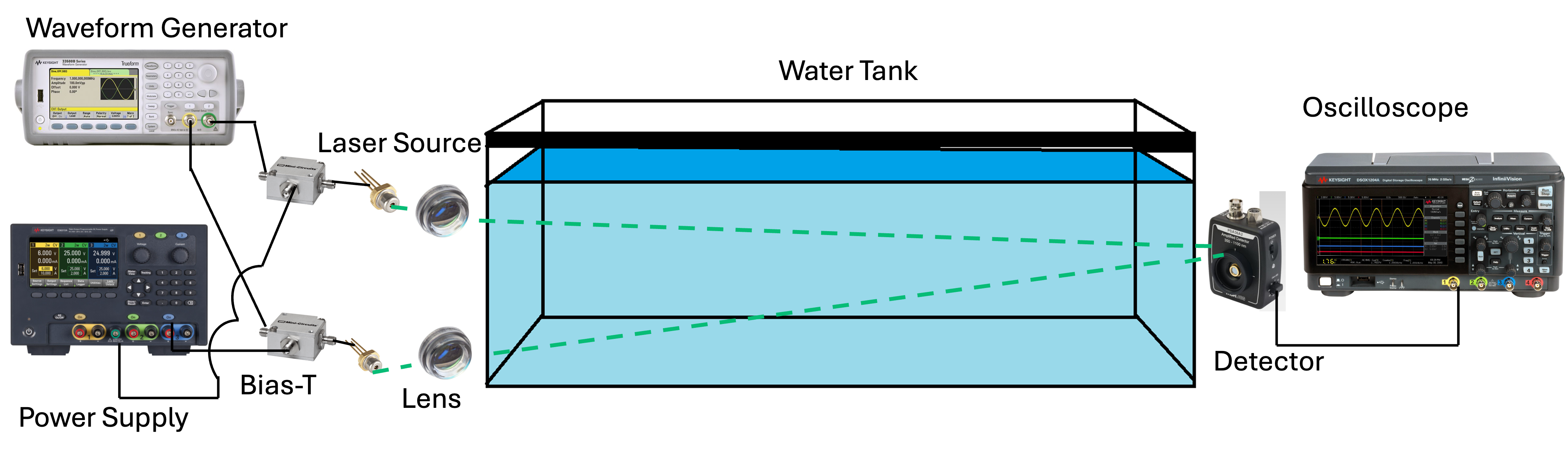}
%\vspace{-2mm}
    \caption{Experimental laboratory design and setup to replicate UOWC for testing CDMA in the tank.}
    \label{fig:testbed-0}
\end{figure*}

\begin{figure*}[!t]
    \centering
\includegraphics[width=\linewidth]{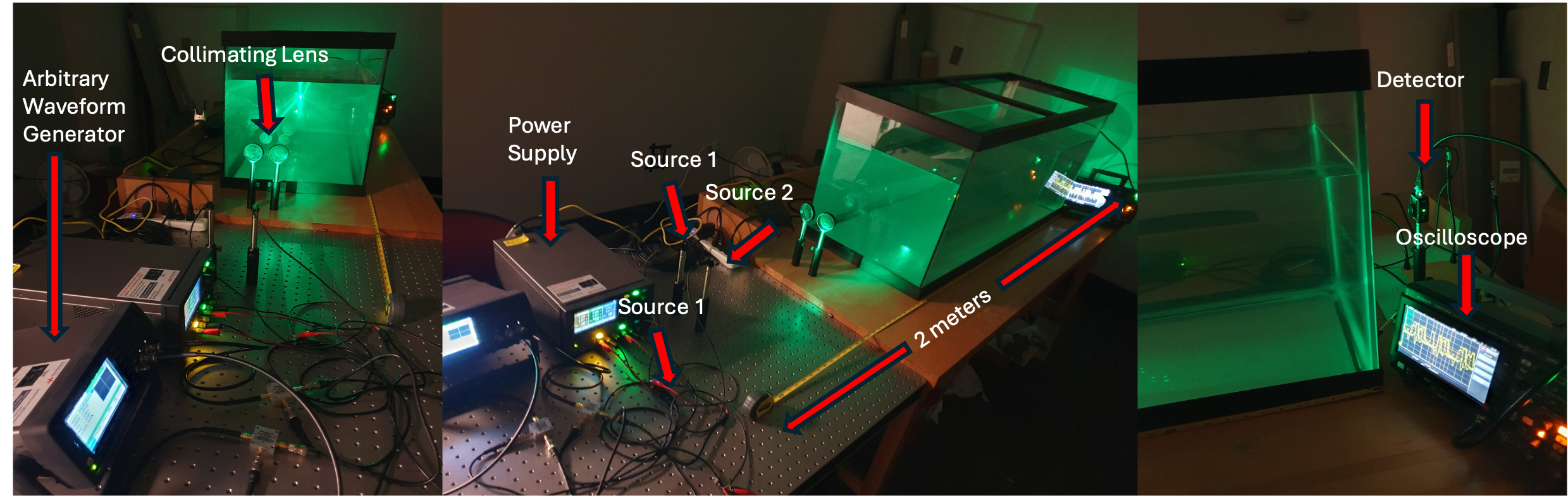}
    \caption{Experimental testbed for UOWC using chaotic-based CDMA, showing (left) the transmitter setup with a green laser diode and convex lens; 
    (middle) water tank experimental testing; and (right) the receiver setup with a concave lens, APD photodiode, and MATLAB interface for real-time data processing and analysis.}
    \label{fig:testbed}
\end{figure*}
\begin{figure*}[!t]
    \centering
\includegraphics[width=0.45\linewidth]{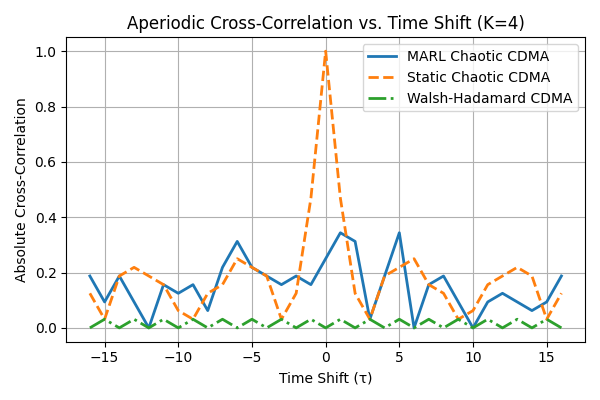}
\includegraphics[width=0.45\linewidth]{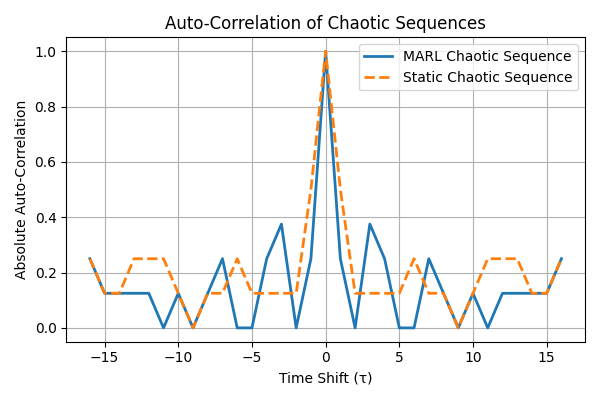}
    \caption{(a) Cross-correlation vs. time shift for MARL chaotic CDMA, static chaotic CDMA, and Walsh-Hadamard CDMA (K=4); (b) Auto-correlation of MARL and static chaotic sequences.}
    \label{fig:new_results}
\end{figure*}

\begin{figure*}[!t]
    \centering
\includegraphics[width=0.325\linewidth]{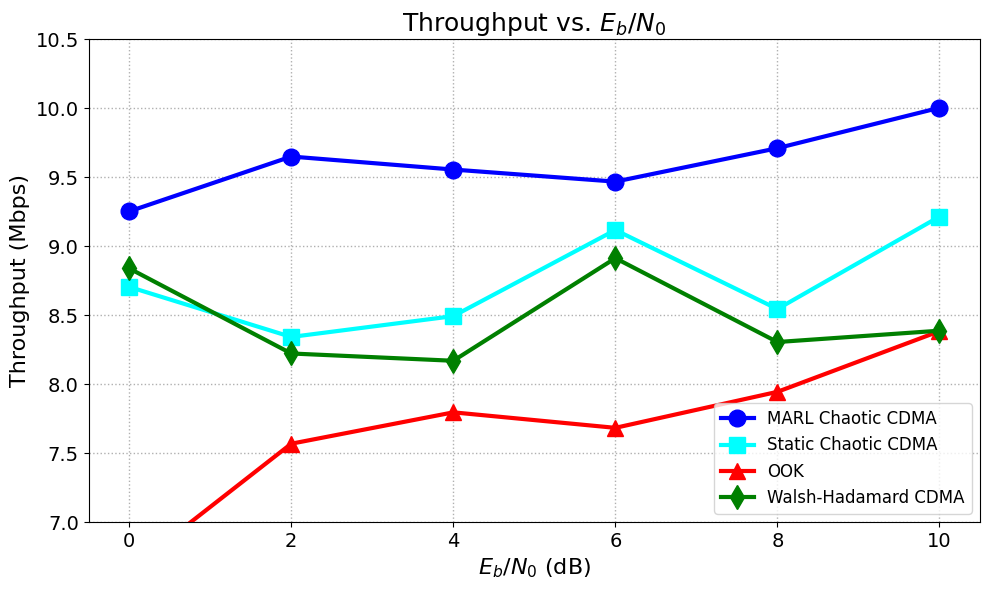}
\includegraphics[width=0.325\linewidth]{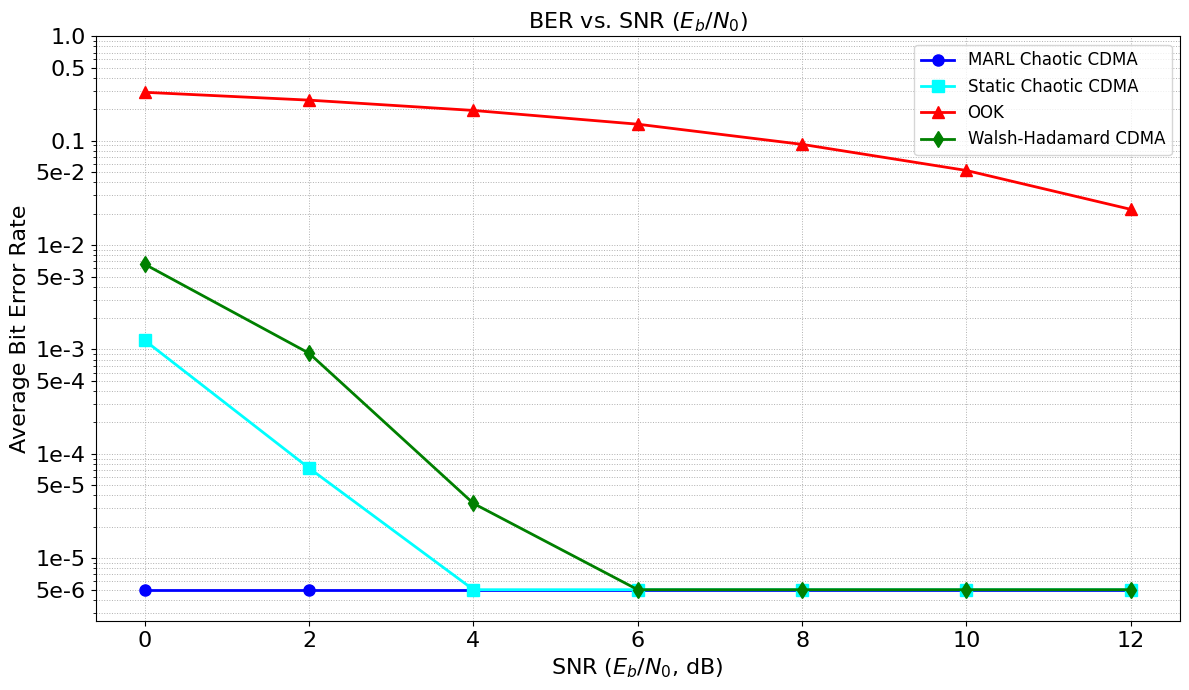}
\includegraphics[width=0.325\linewidth]{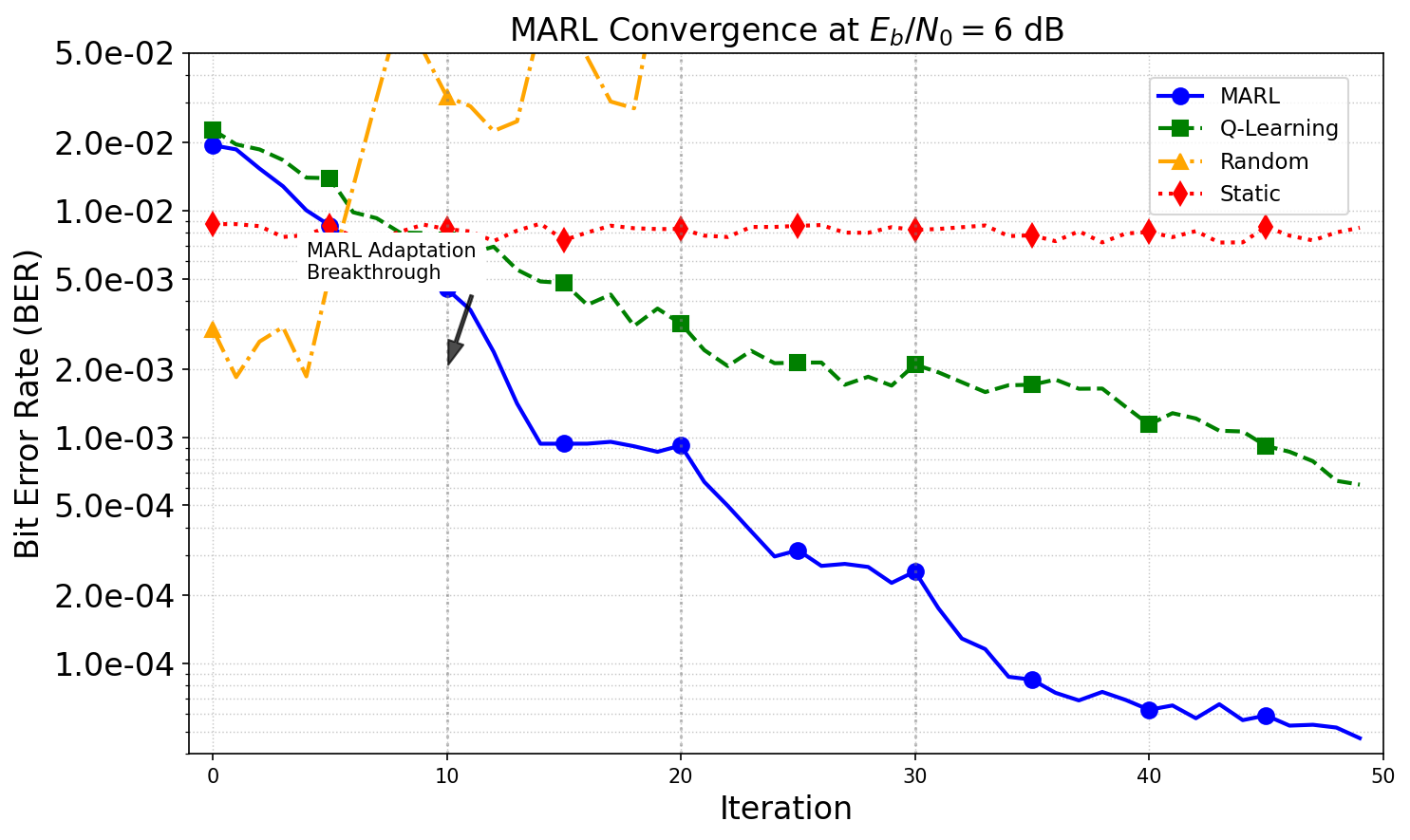}
    \caption{(a) Throughput vs. SNR  for MARL chaotic CDMA, static chaotic CDMA, Walsh-Hadamard CDMA, and OOK in a multipath-affected UOWC environment; (b) BER vs. SNR for four different modulation schemes. The MARL-based scheme achieves the highest throughput across the entire SNR range; and (c) Convergence performance of different algorithms for chaotic CDMA at $E_b/N_0 = 6$ dB.}
    \label{fig:curves}
\end{figure*}
%We conducted both experimental tests and extensive simulations using real-world datasets to evaluate the performance of the proposed MARL-based chaotic CDMA scheme. The following section details the experimental testbed, simulation setup, and analysis of the results.

\textbf{Experimental Testbed:} The experimental testbed was designed to replicate UOWC conditions within a controlled laboratory setting, as shown in Figs.~\ref{fig:testbed-0} and~\ref{fig:testbed}. The transmitter employs a Keysight 33500B waveform generator for signal modulation, paired with a Thorlabs PL520 green laser diode operating at a 520 nm wavelength. A convex lens focuses the laser beam to optimize transmission through the water medium. The channel consists of a 2-meter-long water tank filled with tap and turbid water, with glass walls approximately 1 cm thick. Controlled turbidity levels were introduced using scattering particles to simulate realistic underwater environments. At the receiver, a concave lens captures the scattered optical signal, which is detected by a Thorlabs PDA36A2 silicon avalanche photodiode (APD) for high-sensitivity detection. The electrical signal is digitized using a Keysight DSOX 1204A oscilloscope and processed in MATLAB, where all signal processing tasks, including chaotic sequence generation, data encoding/decoding, modulation/demodulation, and performance metric calculations, are performed. 
%This setup enables real-time evaluation of BER and throughput under varying conditions, such as different turbidity levels and mechanically induced turbulence via stirring.

\textbf{Performance Analysis:}
Figure~\ref{fig:new_results}(a) shows the aperiodic cross-correlation versus time shift for MARL chaotic CDMA, static chaotic CDMA, and Walsh-Hadamard CDMA with 4 users. The MARL scheme (blue) maintains consistently low cross-correlation across shifts, typically below 0.3, demonstrating its effectiveness in minimizing inter-user interference even under asynchronous conditions caused by underwater propagation delays. In contrast, the static chaotic CDMA (orange dashed) exhibits higher peaks, up to 0.8 at zero shift, indicating poorer orthogonality without adaptation. Walsh-Hadamard CDMA (green dash-dot) achieves near-zero cross-correlation at zero shift due to its orthogonal design but shows slight elevations at non-zero shifts, highlighting its vulnerability to timing misalignments in multipath channels. 
%Static chaotic CDMA (orange dashed) degrades rapidly, reaching BER above 0.45 with more users due to fixed sequences leading to higher interference. Walsh-Hadamard CDMA (green dash-dot) performs best at low user counts with BER near 0.25 but plateaus and worsens slightly beyond 6 users, limited by its finite code set and strict synchronization requirements. 
Figure~\ref{fig:new_results}(b) depicts the aperiodic auto-correlation of MARL and static chaotic sequences versus time shift. Both exhibit a sharp peak of 1 at zero shift, essential for reliable synchronization, and drop to low values (below 0.4) at non-zero shifts, confirming desirable thumbtack-like properties for CDMA despreading. The MARL sequence (blue) shows slightly lower sidelobes compared to the static one (orange dashed), reflecting improved spectral characteristics from adaptive parameter tuning, which enhances resistance to ISI in turbid underwater environments.
The performance of the proposed MARL chaotic CDMA scheme is illustrated in Fig.~\ref{fig:curves}(a), which shows the simulated throughput versus SNR in a multipath-affected UOWC environment with moderate turbidity (10 NTU) and weak turbulence. The MARL-based scheme consistently achieves the highest throughput across all SNR levels. At an SNR of 10 dB, a 25\% improvement is observable over static chaotic CDMA and a 60\% gain over OOK. This superior performance results from the dynamic adaptation of chaotic parameters, which minimizes cross-correlation and mitigates time-varying channel distortions. In contrast, static chaotic CDMA and Walsh-Hadamard CDMA suffer from sequence misalignment, reducing their effectiveness in turbulent conditions. Fig.~\ref{fig:curves}(b) presents BER versus SNR curves, demonstrating the robustness of the MARL approach. The proposed method achieves a BER order of $10^{-6}$, significantly lower than static chaotic CDMA, $5 \times 10^{-3}$ for Walsh-Hadamard CDMA, and $10^{-2}$ for OOK. The four-order-of-magnitude improvement over OOK at low SNRs, highlights the effectiveness of chaotic spreading in combating inter-symbol interference and multipath fading. The Walsh-Hadamard scheme's performance degrades in asynchronous scenarios due to propagation delays disrupting code orthogonality, leading to increased multi-user interference. The convergence behavior of the MARL algorithm is shown in Fig.~\ref{fig:curves}(c) at a fixed $E_b/N_0 = 6$ dB, compared to baseline methods like Q-learning and random search. The MARL approach exhibits rapid convergence, with notable breakthroughs at iterations 10, 20, and 30, where agents identify optimal parameter configurations, reducing BER below $5 \times 10^{-5}$. These breakthroughs reflect coordinated adjustments in bifurcation parameters and initial conditions, effectively minimizing cross-correlation and BER. However, Q-learning converges more slowly to a higher BER of $10^{-3}$, and random search plateaus at $10^{-2}$, highlighting the advantage of multi-agent collaboration in navigating the complex parameter space.  Overall, the experimental and simulation results validate the proposed MARL chaotic CDMA scheme’s superior throughput, lower error rates, and adaptability, making it a robust solution for multi-user UOWC in challenging underwater environments.

\section{Conclusion and Future Work} \label{sec:conc}

This paper presented a novel intelligent chaotic code-division multiple access (CDMA) scheme for underwater optical wireless communication (UOWC). The proposed approach leveraged chaotic sequences generated by the logistic map, combined with a multi-agent reinforcement learning (MARL) framework, to dynamically adapt chaotic parameters based on real-time channel conditions.
The key contributions of this work included the development of an adaptive chaotic sequence generation mechanism that maintains low cross-correlation properties even under asynchronous conditions typical of underwater channels, the implementation of a distributed MARL architecture that enabled coordinated parameter optimization across multiple users without centralized control, and the demonstration of relaxed synchronization requirements compared to traditional orthogonal codes, making the system more scalable for distributed underwater networks.
The experimental and simulation results demonstrated the superior performance of the MARL chaotic CDMA scheme across multiple performance metrics. The MARL framework enabled rapid convergence with notable performance improvement, effectively minimizing cross-correlation between users while maintaining robust performance in multi-user scenarios with up to 8 concurrent users.
%Future research directions present several promising avenues for enhancement. 
More experiments in various underwater settings are ongoing and will be published in future papers. 
%The integration of advanced deep learning architectures, such as graph neural networks or transformer-based models, could potentially improve the MARL framework's ability to capture complex channel dynamics and inter-user dependencies. 
The extension to multi-dimensional chaotic systems will provide additional degrees of freedom for sequence optimization and further improve cross-correlation properties. Investigation of hybrid modulation schemes that combine chaotic CDMA with advanced coding techniques will yield additional performance gains, and it will be published in our future work.

\ifCLASSOPTIONcaptionsoff
  \newpage
\fi

\textbf{Acknowledgment:} The authors would like to thank Bryce Lanese, a master's degree graduate of Cleveland State University, for his contributions to the experimental setup.

\bibliographystyle{IEEEtran}
\bibliography{ref}

\end{document}